\documentclass[prb,twocolumn,showpacs,superscriptaddress]{revtex4-1}
\usepackage[latin1]{inputenc}
\usepackage{graphicx}
\usepackage{amssymb}
\usepackage{amsmath}
\usepackage{xspace}
\usepackage{dcolumn}
\usepackage{bm}
\usepackage{color}
\usepackage[extra]{tipa}

\newfont{\tensy}{cmsy10}
\newcommand{\chem}[1]{{$\fontdimen16\tensy=3.0pt
    \fontdimen17\tensy=3.0pt \mathrm{#1}$}}

\newcommand{\ie}[0]{i.e.\@\xspace}
\newcommand{\eg}[0]{e.g.\@\xspace}
\newcommand{\etal}[0]{et al.\@\xspace}

\newcommand{\rmi}{\text{i}}

\newcommand{\on}{\hat{n}}

\newcommand{\oh}{\mbox{$\frac{1}{2}$}}

\newcommand{\om}[0]{\omega}

\newcommand{\nag}{{\phantom{\dag}}}

\newcommand{\las}[0]{\langle}
\newcommand{\ras}[0]{\rangle}
\newcommand{\llas}[0]{\langle\langle}
\newcommand{\rras}[0]{\rangle\rangle}
\newcommand{\la}[0]{\left\las}
\newcommand{\ra}[0]{\right\ras}

\newcommand{\braket}[2]{\la#1\left.\right|#2\ra}

\renewcommand{\tilde}[1]{\widetilde{#1}}

\newcommand{\cdag}{{c}^\dagger}
\newcommand{\cnod}{{c}^{\phantom{\dagger}}}

\begin{document}


\title{%
Phases of correlated spinless fermions on the honeycomb lattice
}

\author{Maria Daghofer}
\affiliation{%
Institut f\"ur Theoretische Festk\"orperphysik, IFW Dresden, 01171 Dresden, Germany}

\author{Martin Hohenadler}
\affiliation{%
\mbox{Institut f\"ur Theoretische Physik und Astrophysik, Universit\"at W\"urzburg,
97074 W\"urzburg, Germany}}

\begin{abstract}  
  We use exact diagonalization and cluster perturbation theory to
  address the role of strong interactions and quantum fluctuations for
  spinless fermions on the honeycomb lattice. We find quantum fluctuations to
  be very pronounced both at weak and strong interactions. A weak
  second-neighbor Coulomb repulsion $V_2$ induces a tendency toward
  an interaction-generated quantum anomalous Hall phase, as borne out
  in mean-field theory. However, quantum fluctuations prevent the formation
  of a stable quantum Hall phase before the onset of the charge-modulated
  phase predicted at large $V_2$ by mean-field theory. Consequently, the
  system undergoes a direct transition from the semimetal to the
  charge-modulated phase. For the latter, charge fluctuations also play a key
  role. While the phase, which is related to pinball liquids, is stabilized
  by the repulsion $V_2$, the energy of its low-lying charge excitations
  scales with the electronic hopping $t$, as in a band insulator.
\end{abstract} 

\date{\today}

 
\maketitle

\section{Introduction}

Interest in topological states of matter has been boosted by the discovery of
topological insulators and superconductors.\cite{HaKa10,RevModPhys.83.1057}
The quantum spin Hall insulator, a novel topological state of matter with a
$Z_2$ topological invariant and helical edge states\cite{KaMe05a,KaMe05b} has
been observed experimentally in HgTe quantum well
structures.\cite{BeHuZh06,Koenig07} Since then, research has
broadened substantially, and now includes three-dimensional topological
insulators and superconductors,\cite{RevModPhys.83.1057} fractional Chern
insulators,\cite{Pa.Ro.So.2013,2013arXiv1308.0343B} and symmetry-protected
topological
phases.\cite{PhysRevB.82.155138,arXiv:1106.4772,PhysRevB.84.235141,Chen2012,arXiv:1209.4399,arXiv:1212.1726}

Topological insulators and Chern insulators typically arise from complex
hopping terms related to spin-orbit coupling\cite{KaMe05a,KaMe05b} or to a
periodic vector potential.\cite{Haldane98} By now, several noninteracting
models are known that support quantum Hall and quantum spin Hall
phases.\cite{HoAsreview2013} In such settings, electron-electron interactions
play a minor role as a result of the bulk band gap, and the states closely
resemble band insulators.  Sufficiently strong interactions can drive
transitions to nontopological phases with
magnetic\cite{RaHu10,Hohenadler10,Zh.Wu.Zh.11} or charge-density-wave
order.\cite{VaSuRi10,Va.Su.Ri.Ga.11,Wa.Sh.Zh.Wa.Da.Xi.10} An interesting
interaction-driven transition from a quantum spin Hall (QSH) phase to
a phase with fractional excitations and topological order is the so-called
QSH* phase found in a mean-field treatment of a model for
\chem{Na_2IrO_3}.\cite{PhysRevLett.108.046401} The interplay of topological
band structures and electronic interactions has been studied extensively, see
Ref.~\onlinecite{HoAsreview2013} for a review.

Conversely, electronic correlations can also give rise to topological states.
Topological Mott insulators,\cite{RaQiHo08} or interaction-generated
topological insulators, are a particularly interesting concept. Raghu
\etal\cite{RaQiHo08} presented a scenario where quantum (spin) Hall states
arise purely from electronic interactions that give rise to spontaneously
generated, complex bond order
parameters.\cite{RaQiHo08,PhysRevLett.93.036403} Such a correlation-driven
route to topological states would abandon the requirement of strong intrinsic
spin-orbit coupling and could thus significantly extend the class of
topologically nontrivial materials.  In a more general context,
correlations have been shown to stabilize topologically nontrivial bands in
double-exchange models on the kagome and triangular
lattices,~\cite{PhysRevB.62.R6065,Martin08} where the topological character
is supported by the coupling to localized spins. In particular, it has
been shown that Haldane's scenario of a transition from bands featuring Dirac
cones to bands with a topologically nontrivial gap can be observed in the
Kondo-lattice model on the checkerboard lattice.~\cite{Venderbos:2012hv}
The ordering of complex orbitals can also lead to topological
insulators.\cite{PhysRevB.84.201104,Ru.Fi.11,PhysRevB.85.245131,PhysRevB.84.241103}

For the spinless model considered in Ref.~\onlinecite{RaQiHo08}, the
existence of a topological phase, namely a quantum anomalous Hall (QAH)
state, has been confirmed by more elaborate mean-field
approximations.\cite{PhysRevB.81.085105,GrushinII} More generally, again
using mean-field and renormalization group methods, interaction-generated
topological states have been shown to arise in kagome, checkerboard, or
decorated honeycomb
lattices,\cite{PhysRevLett.103.046811,PhysRevB.82.045102,PhysRevB.82.075125}
in a $\pi$-flux square lattice model,\cite{PhysRevB.81.085105} as well as in
three dimensions.\cite{PhysRevB.79.245331}  In contrast to checkerboard and kagome
lattices, the Dirac points in the $\pi$-flux and honeycomb model are
associated with a vanishing density of states at the Fermi level. Consequently,
transitions to symmetry-broken phases may not be correctly captured by a
weak-coupling approach.  The variety and fascinating properties of these
novel phases make it desirable to go beyond a weak coupling description.
Indeed, recent exact diagonalization results\cite{Ji.Gu.Ch.Sh.Fe.13} for the
$\pi$-flux square lattice model have not confirmed mean-field predictions of
an interaction-generated QAH phase.\cite{PhysRevB.81.085105}

Here, we use exact diagonalization to study the spinless model
first considered in Ref.~\onlinecite{RaQiHo08}. Thereby, we fully take into
account quantum fluctuations which are expected to be strong given the low
coordination number of the honeycomb lattice. Most importantly, our results
imply that the interaction-generated topological mean-field state is unstable
with respect to fluctuations, and that the gapped ground state is not
adiabatically connected to the QAH state of the Haldane model. 
However, we demonstrate that for small $V_2$, the model has a
tendency toward an interaction-generated QAH state. In addition,
we provide new insights into the charge-ordered phase that exists for
strong next-nearest-neighbor repulsion.\cite{GrushinII} The organization of
this paper is as follows. In Sec.~\ref{sec:model}, we define the model
considered. Our results are discussed in Sec.~\ref{sec:results}, and we
conclude in Sec.~\ref{sec:conclusions}.

\section{Model}\label{sec:model}

Following Ref.~\onlinecite{RaQiHo08}, we consider a model of interacting,
spinless fermions described by the Hamiltonian
\begin{eqnarray}\label{eq:modelH1}
  \hat{H}_1
  = 
  \hat{H}_0 
  + V_1 \sum_{\las i j\ras} \on_i \on_j
  + V_2 \sum_{\llas i j\rras}  \on_i\on_j
  \,
  .
\end{eqnarray}
The first term,  $\hat{H}_0= -t \sum_{\las ij \ras} (c^\dag_i c^\nag_j + c^\dag_j c^\nag_i)$,
describes nearest-neighbor (NN) hopping on the honeycomb lattice. The second term accounts for a
repulsion between fermions on NN sites, whereas the third term describes a
repulsion between next-nearest-neighbor (NNN) sites (\ie, sites on the same
sublattice). The indices $i,j$ number lattice sites, and $L$ denotes the
total number of sites. Throughout this paper, we consider a half-filled band
with one fermion per unit cell and $\las n_i \ras=1/2$.

Hamiltonian~(\ref{eq:modelH1}) was previously studied at the mean-field
level.\cite{RaQiHo08,PhysRevB.81.085105,GrushinII} These works reported
a QAH state with chiral edge states and a nonzero Chern index. This phase is
characterized by a complex bond order parameter\cite{RaQiHo08} $\chi_{ij}=\chi^*_{ji}=\las
c^\dag_i c^\nag_j\ras$ that mimics the complex hopping term
of the Haldane model\cite{Haldane98} and breaks time-reversal symmetry.
The QAH state is driven by $V_2$ and, according to mean-field
theory, most stable for $V_1=0$.\cite{RaQiHo08,PhysRevB.81.085105,GrushinII}
For $V_1=0$, Refs.~\onlinecite{RaQiHo08,PhysRevB.81.085105} found a
semimetal (SM) and a QAH phase. The SM is stable up to a finite critical value of
$V_2$ because of the vanishing density of states at the Fermi level. Using
a more elaborate mean-field ansatz, Grushin \etal\cite{GrushinII} obtained an additional, charge-modulated (CM)
insulating phase at large $V_2/t$ that restricts the QAH phase to a finite region $1.5\lesssim
V_2/t \lesssim 2.5$.  For $V_1>0$, a charge-density-wave phase with broken
inversion symmetry\cite{RaQiHo08,PhysRevB.81.085105,GrushinII} (for
$V_1>V_2$), as well as a Kekul\'{e} ordered phase with broken translational
invariance (for $V_1\sim V_2$) were
found.\cite{PhysRevLett.98.186809,PhysRevB.81.085105,GrushinII,PhysRevLett.98.186809}
The low-energy field theory of interacting spinless fermions on the honeycomb
lattice is discussed in Ref.~\onlinecite{PhysRevB.79.085116}.

\section{Results}\label{sec:results}

According to mean-field theory,\cite{RaQiHo08,PhysRevB.81.085105,GrushinII}
the QAH phase is stabilized by $V_2$, and is therefore most extended in
parameter space for $V_1=0$. Therefore, and to simplify the analysis, we
focus on the case $V_1=0$, although some results for nonzero $V_1$ will also be
presented.

The exact diagonalization results presented below have been obtained
on clusters with 18, 24, and 30 sites, respectively. Since the Dirac points $\pm
K$ define the low-energy physics of the noninteracting system ($V_1=V_2=0$),
and also correspond to the ordering wavevector of the charge order driven by
large values of $V_2$ (see below), we have chosen clusters for which
$\bm{q}=\pm K$ are allowed momenta. In the notation of Ref.~\onlinecite{VaSuRi10},
the clusters used here correspond to 18A and as well as 24A; results for a
small number of parameter sets were also obtained using 30A. We have verified
that our findings are unchanged when using clusters 24C and 24D.

\subsection{Phase diagram for $V_1=0$}\label{sec:V0}

The QAH state found in mean-field
theory\cite{RaQiHo08,PhysRevB.81.085105,GrushinII} is identical to the QAH
ground state of the noninteracting Haldane model,\cite{Haldane98} and hence
characterized by a Chern number $C=\pm1$. To prove the existence of this
phase numerically, it is not sufficient to simply calculate the Chern index
for the model~(\ref{eq:modelH1}). The reason is that there exist two possible
bond-order patterns which differ by an overall sign, and describe
Chern insulators with $C=1$ and $C=-1$, respectively. When the ground state
of a finite cluster is determined by exact diagonalization, it can be
expected to be a linear combination of these two states, and hence to have a
vanishing Chern index. Finally, the accessible system sizes are not sufficient
to carry out a finite-size extrapolation to the thermodynamic limit to reveal
a symmetry breaking. Given these complications, a different route has to be chosen.

To identify a possible QAH state driven by the interaction $V_2$, we here study a
superposition\cite{PhysRevLett.109.246805} of the Hamiltonian of interest, namely Eq.~(\ref{eq:modelH1}),
and a Hamiltonian known to have the QAH ground state predicted by mean-field
theory. The mean-field QAH state of
Eq.~(\ref{eq:modelH1}), first reported in Ref.~\onlinecite{RaQiHo08},
is identical to the QAH state of the Haldane Hamiltonian\cite{Haldane98}
\begin{eqnarray}\label{eq:haldane}
  \hat{H}_2
  = 
  \hat{H}_0
  &-t_2& \sum_{\llas ij \rras} ( e^{\rmi \phi_{ij}} c^\dag_i c^\nag_j + e^{-\rmi \phi_{ij}} c^\dag_j c^\nag_i) \,,
\end{eqnarray}
for the choice of phase $\phi_{ij}=\pm\pi/2$; the sign depends on the
direction of the bond $\llas ij\rras$ and the sublattice. It  arises from
periodic magnetic fluxes that sum to zero for each hexagon of the
honeycomb lattice.\cite{Haldane98} The hopping term $\hat{H}_0$ is
identical to Eq.~(\ref{eq:modelH1}). Hence, {\it at the mean-field level}, the
Hamiltonian
\begin{equation}
  \tilde{H}(\lambda)=\lambda H_2 + (1-\lambda) H_1\,, \quad \lambda\in[0,1]\,,
\end{equation}
interpolates between the noninteracting QAH ground state of the Haldane
Hamiltonian $\hat{H}_2$ (\ie, $t_2>0$, $V_2=0$, $\lambda=1$) and the
interaction-generated QAH ground state of $\hat{H}_1$ ($\lambda=0$, and
suitable values of $V_2$). As a function of $\lambda$, it is therefore
possible to adiabatically connect the ground states that exist for
$\lambda=0$ and $\lambda=1$. For $0<\lambda<1$,
$\tilde{H}(\lambda)$ describes interacting fermions on the honeycomb lattice
with additional Haldane hopping $t_2$.  

If the mean-field QAH state is stable, a similar adiabatic connection between
$\lambda=0$ and $\lambda=1$ is expected to exist when $\tilde{H}(\lambda)$ is
solved using exact numerical methods. Starting with $\lambda=0$,
we hence expect a continuous evolution with $\lambda$ if the ground
state of $\hat{H}_1$ is indeed a QAH state. In particular, switching on $t_2$
should reinforce a potential QAH ground state of $\hat{H}_1$.  Conversely,
a discontinuous evolution (\eg, a phase transition) as a function of
$\lambda$ would imply that the state at $\lambda=0$ is not the QAH state
predicted by mean-field theory.

Here, we calculate the quantum fidelity $F=\braket{\phi_0(V_2+\delta
  V_2)}{\phi_0(V_2)}$, corresponding to the overlap of the ground states of
Hamiltonian~(\ref{eq:modelH1}) for $V_2$ and $V_2+\delta V_2$,
respectively, with all other parameters unchanged. The fidelity permits us to detect transitions between different
phases without making assumptions regarding order
parameters.\cite{PhysRevE.74.031123,PhysRevE.76.022101} Moreover, it is
particularly suitable to detect transitions between topologically trivial and
nontrivial insulators,\footnote{More precisely: Between phases whose Chern
  numbers differ by an odd number.} because such transitions involve a level
crossing even on finite clusters.\cite{VaSuRi10,Va.Su.Ri.Ga.11} In contrast,
continuous symmetry-breaking transitions appear as gradual changes on finite clusters,
and are therefore often difficult to identify.

\begin{figure} 
  \includegraphics[width=0.425\textwidth]{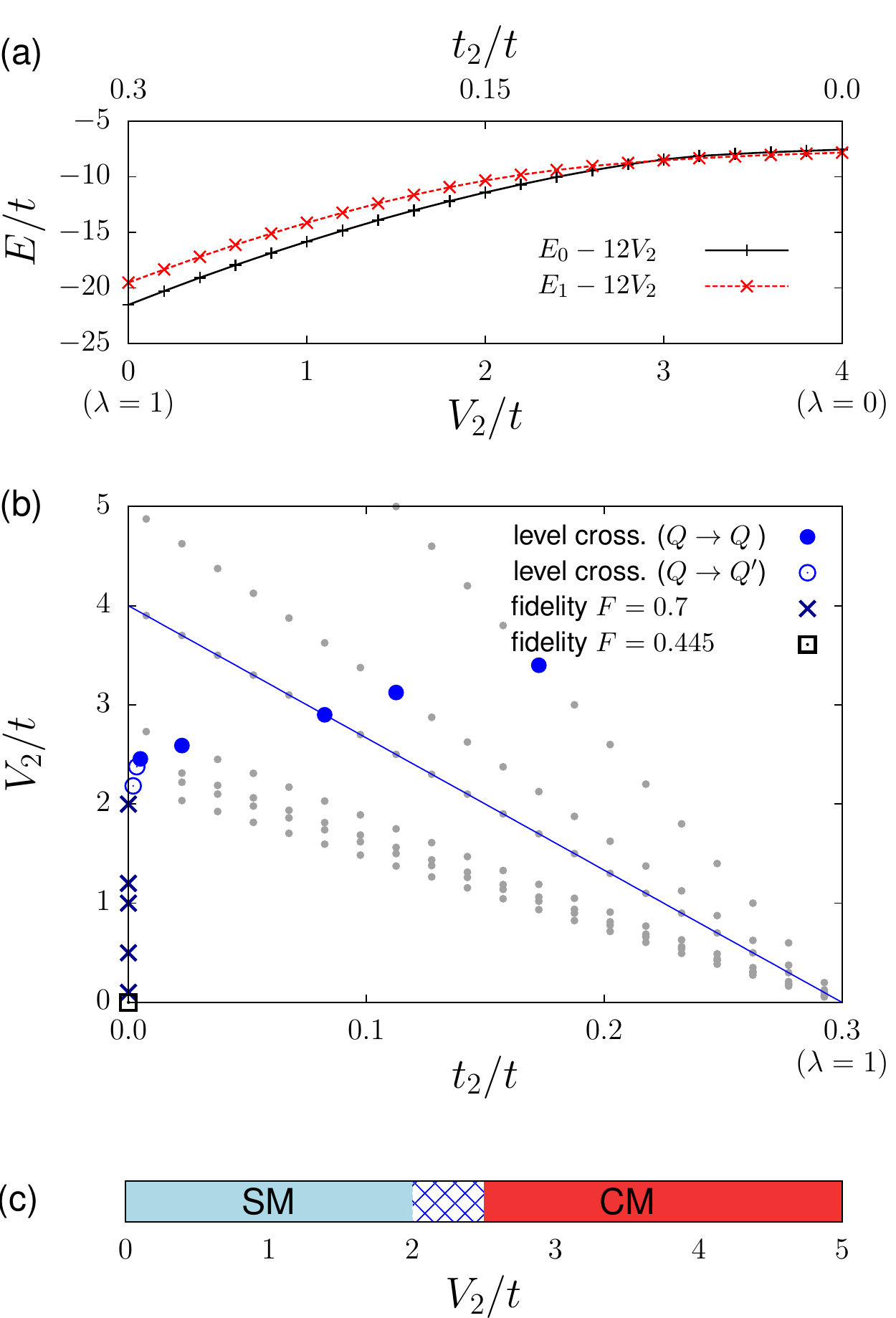}
  \caption{\label{fig:trans} (Color online) (a) The two lowest energy levels
    in the ground state momentum sector along the path from
    $(t_2=0.3,V_2=0)$ to $(t_2=0,V_2=4)$, indicated by the solid line in
    (b).  (b) Moving along the parameter trajectories indicated by the small
    dots from the QAH state of the Haldane model~(\ref{eq:haldane}) at
    $t_2=0.3t, V_2=0$ ($\lambda=1$) toward the model~(\ref{eq:modelH1}) with
    $t_2=0$ and different $V_2$ ($\lambda=0$), we find level crossings at the
    points indicated by large circles. Large open (filled) circles correspond to
    level crossings with (without) a change of the ground-state momentum
    sector. Results in (a) and (b) are for $V_1=0$. (c) Phase diagram of
    Hamiltonian~(\ref{eq:modelH1}) with  $V_1=0$. The hatched region reflects
    the uncertainty regarding the  critical point due to finite-size
    effects. All results were obtained from
    exact diagonalization of $\tilde{H}$ using a 24-site cluster.}
\end{figure}

We take the mean-field phase diagram as a starting point, and distinguish
three regimes. For sufficiently large $V_2$ (the mean-field
prediction is $V_2\gtrsim2.5t$), the gapped CM phase is
expected.\cite{GrushinII} For smaller $V_2$ ($1.5 \lesssim V_2/t\lesssim
2.5$, according to Ref.~\onlinecite{GrushinII}), the QAH state exists, and
for $V_2\lesssim 1.5t$, mean-field theory finds the SM phase.

\begin{figure}[t]
  \includegraphics[width=0.425\textwidth,clip]{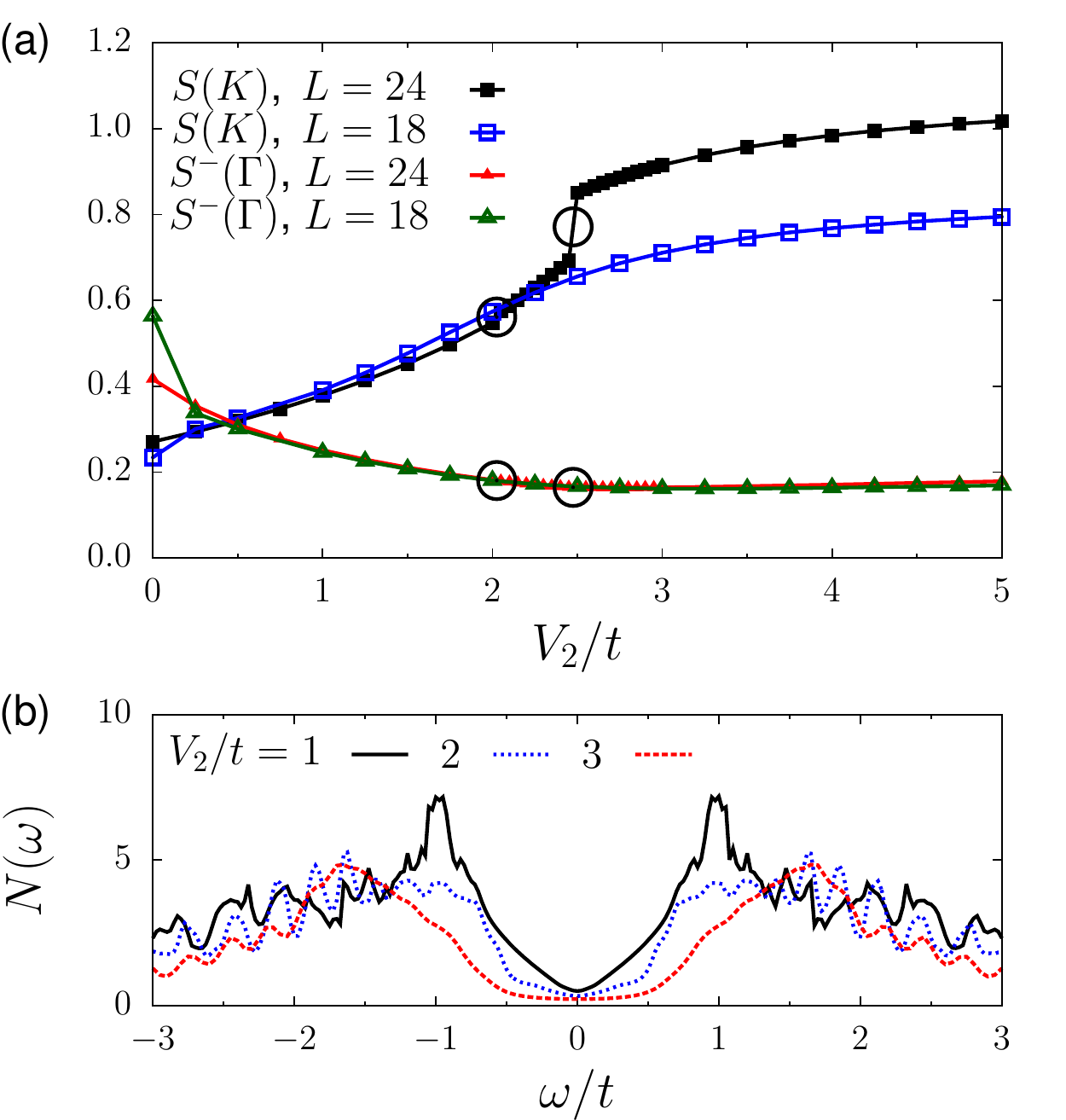}
  \caption{\label{fig:phaseszerot2}
    (Color online) (a) Charge structure factors $S(K)$ and $S^-(\Gamma)$, see
    Eq.~(\ref{eq:Nk}),  from exact diagonalization for two different cluster
    sizes ($L=18$ and $L=24$). The open circles at $V_2\approx 2t$ and $V_2\approx 2.5t$
    indicate the change of the ground state momentum sector observed for $L=24$.
    (b) Density of states for different values of $V_2/t$ obtained
    from exact diagonalization with twisted boundary conditions and $L=24$.
    All results are for $V_1=0$.
  }
\end{figure}

We first consider the CM region.  In Fig.~\ref{fig:trans}(a), we show the
evolution of the two lowest energy levels of $\tilde{H}(\lambda)$ along a
path from $(t_2,V_2)=(0.3t,0)$ ($\lambda=1$) to $(t_2,V_2)=(0,4t)$
($\lambda=0$), as indicated by the solid line in
Fig.~\ref{fig:trans}(b). Whereas the point $\lambda=1$ lies in the
well-established QAH phase of the Haldane model, the point $\lambda=0$ has a
sufficiently large $V_2$ to fall into the CM phase.\cite{GrushinII} [The
existence of charge order will be demonstrated below, see
Fig.~\ref{fig:phaseszerot2}(a).] Since $\lambda=1$ corresponds to the Haldane
model, the initially lower-lying level (solid line) in Fig.~\ref{fig:trans}(a) can
be identified with the QAH state with Chern number $C=1$. We find
that switching on $V_2$ in the Haldane model does not immediately destroy
the QAH state, as can be expected for a gapped phase. However, at a critical
value $V_2\approx 2.9t$ (and $t_2\approx 0.08t$) we observe a level crossing
within the same momentum sector, and a vanishing of the fidelity. This level
crossing, signaling a quantum phase transition to a topologically distinct
state, reveals that the QAH state at $t_2 =0.3t$, $V_2=0$ is different from
the gapped ground state at $t_2=0$, $V_2=4t$, in accordance with the
mean-field theory prediction of a CM phase for these parameters.
 
Figure~\ref{fig:trans}(b) shows the level crossings found along similar
paths in the $( t_2,V_2)$ plane, but with end points that have different values
of $V_2$. For values as small as $V_2=2.5t$, we find the same type of level
crossing as illustrated in Fig.~\ref{fig:trans}(a). Moreover, with decreasing
$V_2/t$, the level crossings move toward smaller values of $t_2$, in accordance
with the decrease of the gap of the CM state.

In the regime $2t\lesssim V_2\lesssim 2.5t$, a level crossing occurs at small
but finite values of $t_2$, but between different ground-state momentum
sectors (indicated by open circles). Moreover, the ground state in this
regime is doubly degenerate with momenta $\pm K$, as opposed to the
nondegenerate ground state with momentum $\Gamma$ that exists for $V_2/t$
outside $[2,2.5]$.\footnote{Except for the point $t_2=V_1=V_2=0$, where the Dirac
  semimetal is six-fold degenerate.} This behavior can be understood by
considering the model with $t_2=0$, see Fig.~\ref{fig:phaseszerot2}(a) and
discussion below, for which we observe two changes of the momentum sector as
a function $V_2$ at $V_2\approx 2t$ ($\Gamma\to\pm K$) and $V_2\approx2.5t$
($\pm K\to\Gamma$). We attribute the
existence of this intermediate regime and the momentum changes to finite-size
effects related to the close energetic proximity of excited states with
momentum $\bm{q}=\pm{K}$ to the ground state in the SM and the CM phases. Indeed, the
momentum-changing level crossings are absent on $L=18$ and $L=30$ clusters, and a
similar cluster-dependent intermediate region has been reported for the
interacting Haldane model.\cite{Va.Su.Ri.Ga.11} More importantly, the fact
that a level crossing occurs as a function of $\lambda$ implies that the
ground state of $\hat{H}_1$ is not adiabatically connected to the QAH state of the
Haldane model down to $V_2\approx 2t$. 

Finally, for interactions $V_2\lesssim 2t$, the fidelity
$F=\braket{\phi_0(V_2,t_2)}{\phi_0(V_2,t_2+\delta t_2)}$ with $\delta t_2>0$,
which is very close to 1 for $t_2>0$, decreases to 0.7 (or $1/\sqrt{2}$) for
$t_2=0$; the corresponding parameters are indicated by the crosses in Fig.~\ref{fig:trans}(b).
The fact that any finite $t_2$ significantly modifies the ground state suggests that in this regime, we have
the SM phase which is unstable toward the opening of a topological mass gap
by a finite $t_2$. The same behavior can be observed in the noninteracting
Haldane model.\cite{Haldane98}  We observe the same fidelity over the range $V_2\in
[0,2t]$, which suggests that the SM phase extends at least up to $V_2=2t$. This
value is comparable to the mean-field estimates.\cite{RaQiHo08,PhysRevB.81.085105,GrushinII}

The results of this section, in particular the fact that the gapped parameter
region of Hamiltonian~(\ref{eq:modelH1}) cannot be adiabatically connected to
the QAH state of the Haldane model~(\ref{eq:haldane}), suggest that the
mean-field prediction of a QAH phase is not borne out. Instead, we propose the
$V_1=0$ phase diagram shown in Fig.~\ref{fig:trans}(c), with a direct
transition from the SM to the CM phase at a critical value $V_2\approx
2.5t$. We will see below that this scenario is consistent with the results
for charge structure factors [note the jump of $S(K)$ in Fig.~\ref{fig:phaseszerot2}
near $V_2=2.5t$] and the density of states.

\subsection{Charge order driven by $V_2$} \label{sec:CM_V2}

Given a two-site unit cell, two charge structure factors
$S^{\pm}({\bf q}) $ can be defined for each sublattice momentum ${\bf q}$,
which differ by the relative phase between the contributions of the two
sublattices and can be written as
\begin{align}\label{eq:Nk}
  \hspace*{-1em}
  S^{\pm}({\bf q}) 
  = \frac{1}{L}
  \Big|\sum_{ j} \textrm{e}^{i{\bf q}\cdot{\bf r}_j} 
  \left[ (\hat{n}^A_{j} -\oh) \pm (\hat{n}^B_{j} -\oh) \right]
  | \phi_0\rangle \Big|^2\,.
\end{align}
Here, $\hat{n}^\alpha_{j}$ is the density operator for a site on sublattice
$\alpha$ in unit cell $j$, and
$|\phi_0\rangle$ denotes the many-body ground state. A N\'eel-type charge
order corresponding to a sublattice charge imbalance within the unit cell, as
previously observed for spinless fermions with interaction $V_1$,\cite{RaQiHo08}
is captured by $S^-(\Gamma)$ with $\Gamma=(0,0)$, whereas the charge order
predicted by mean-field theory\cite{GrushinII} for large $V_2$ can be tracked
by $S(K)\equiv \oh [S^+(K)+S^-(K)]$.

Figure~\ref{fig:phaseszerot2} shows results for these structure
factors for 24 and 18-site clusters, obtained for the original
model~(\ref{eq:modelH1}). The N\'eel structure factor $S^-(\Gamma)$ is
quickly suppressed from its noninteracting value with increasing $V_2$,
while $S(K)$ is enhanced. The open circles in Fig.~\ref{fig:phaseszerot2}
indicate where the ground state of the $L=24$ cluster changes momentum, see
discussion above. Whereas $S(K)$ continues to grow in this regime,
$S^-(\Gamma)$ is almost unchanged. This finding suggests that this
intermediate regime is not a different phase, since in that case we would
expect the charge order [\ie, $S(K)$] to be suppressed. Other potential order
parameters that we considered (including bond order) are similarly
unaffected in this parameter region. Moreover, while a finite $t_2$ is
needed to move from this regime to the QAH state, the very small critical
values (e.g., $t_2=0.003t$ for $V_2=2.3t$) are a strong argument against any
gapped intermediate phase.  

\begin{figure}[t]
  \includegraphics[width=0.425\textwidth,clip]{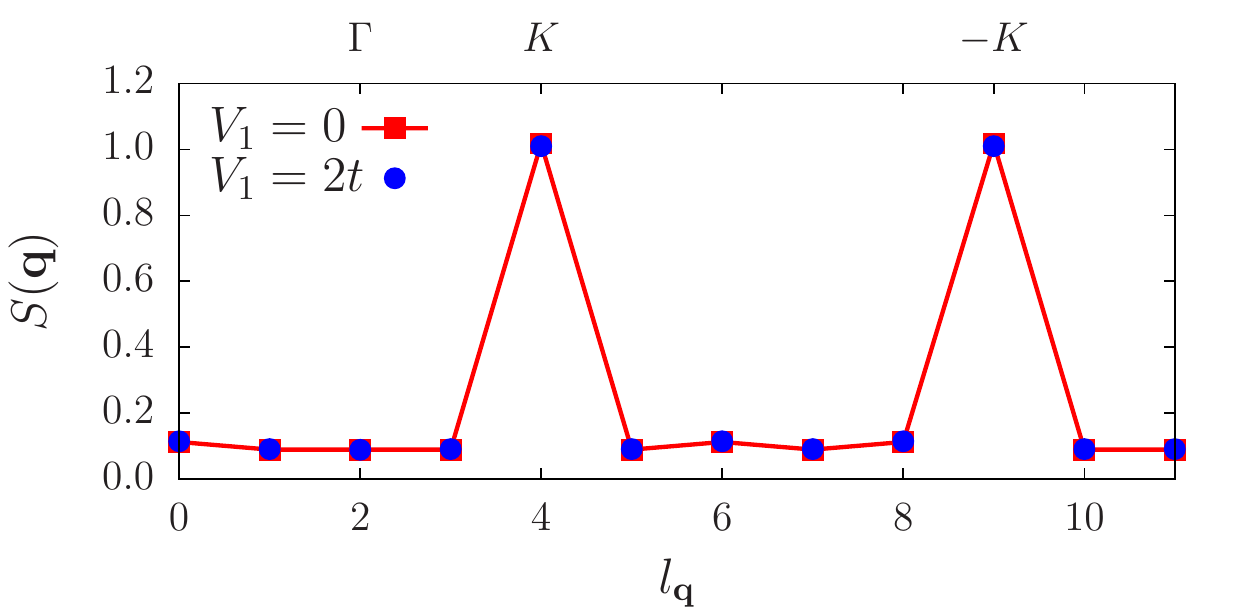}
  \caption{\label{fig:Nk_CM}
    (Color online) Charge structure factor $S(\bm{q})$ for $L=24$
    sites, $t_2=0$, $V_2=5 t$ (deep in the CM phase), and two
    values of the NN Coulomb repulsion $V_1$. Here, $l_{\bf q}$ indexes the
    wavevectors $\bm{q}$, with $l_{\bm{q}}=4$, 9 corresponding to $\bm{q}=\pm K$ and
    $l_{\bm{q}}=2$ corresponding to $\bm{q}=\Gamma$. Since $S^+(\Gamma)=0$,
    $S(\Gamma)$ is proportional to the N\'eel structure factor $S^-(\Gamma)$.
  }
\end{figure}

These results, together with the absence of an
intermediate region where the ground state momentum changes on the $L=18$ and $L=30$
clusters, agree with our previous conclusion, namely a direct transition from the
SM to the CM phase.  Keeping in mind that a finite-size scaling is not
feasible with the accessible cluster sizes, our estimate for the critical
point is $V_2\approx 2.5t$. Figure~\ref{fig:phaseszerot2}(b) shows the density of states obtained for a
24-site cluster. Our use of twisted boundary conditions reduces finite-size
effects by enhancing the resolution in momentum space, but can induce small
spurious gaps as a result of the breaking of translational
symmetry.\cite{PhysRevLett.111.029701} Nevertheless, the results in
Fig.~\ref{fig:phaseszerot2}(b) suggest the existence of a single-particle gap
for $V_2 \gtrsim 2.5 t$, in accordance with the phase diagram shown in
Fig.~\ref{fig:trans}(c).

An insulating, charge-ordered phase at large values of $V_2/t$ (referred to
as the CM phase) was first observed in Ref.~\onlinecite{GrushinII}, and  
a related charge-ordered phase has been reported for spinful fermions in
honeycomb bilayers and trilayers.\cite{PhysRevB.86.155415,PhysRevB.85.235408}
At the mean-field level, the pattern of charge density deviations from half filling
takes the form $+\delta,-\delta,+\delta,+\Delta,-\Delta,-\delta$ (with
$\Delta>\delta$) for consecutive sites of the hexagonal unit
cell.\cite{GrushinII} Within each sublattice, charge is modulated with a
three-site unit cell, corresponding to an ordering wavevector $K$, and charge
is in general unequally distributed between the sublattices (except for
$\Delta=2\delta$). The CM phase is different from the charge-density-wave
phase with a N\'eel-type charge modulation which is driven by large values of
$V_1$.\cite{RaQiHo08,GrushinII}

Our exact diagonalization results shown in Fig.~\ref{fig:phaseszerot2}
suggest a transition from the SM to the CM phase at $V_2\gtrsim 2.5t$.  The
fact that the N\'eel structure factor $S^-(\bm{q})$ remains comparable to
$S(\bm{q})$ at other momenta $\bm{q}\neq \pm K$, see Fig.~\ref{fig:Nk_CM},
provides an argument against charge imbalance between the sublattices. On the
other hand, the N\'eel signal $S^-(\Gamma)$ is not suppressed when we move
deeper into the CM phase (larger $V_2\gg 2t$) either, see
Fig.~\ref{fig:phaseszerot2}(a), as one would expect for the mean-field phase
with $\Delta=2\delta$.

\begin{figure}[t]
  \includegraphics[width=0.425\textwidth,clip]{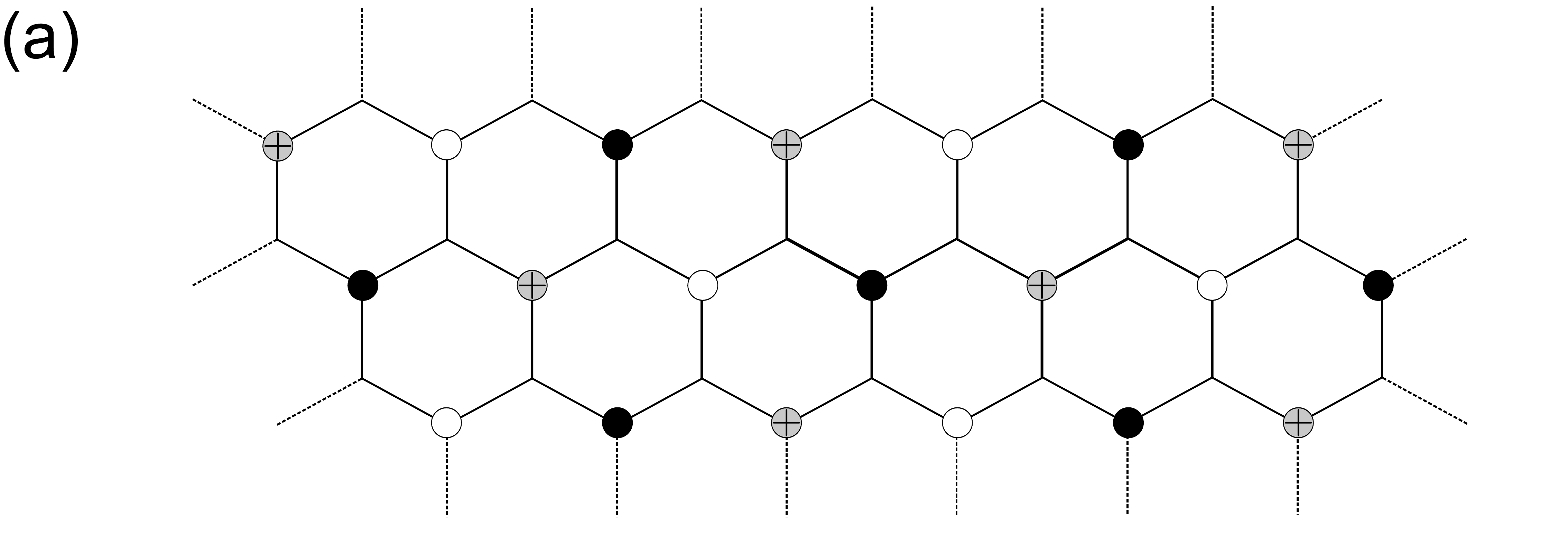}\\
  \includegraphics[width=0.425\textwidth,clip]{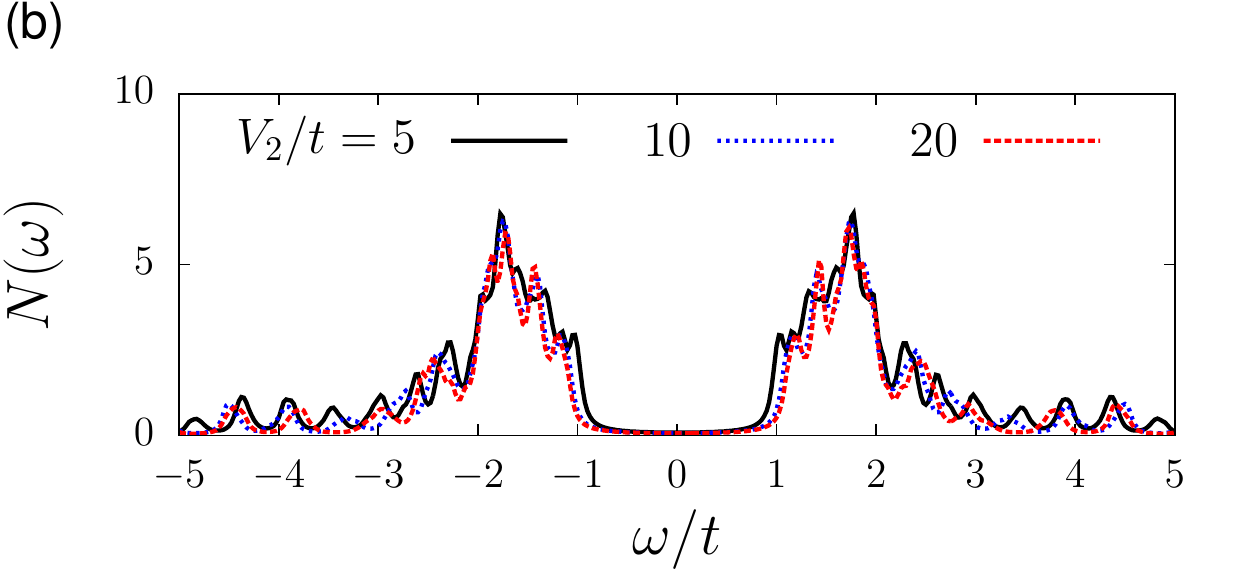}
  \caption{\label{fig:cmphase}
    (Color online) 
    (a) Atomic-limit charge distribution for one of the two
    sublattices. Filled (empty) circles indicate occupied (empty) lattice sites. Any distribution
    of the remaining fermions over the $\oplus$ sites gives the same energy.
    (b)
    Density of states in the CM phase from exact diagonalization with twisted
    boundary conditions and $L=24$.
  }
\end{figure}

A closer analysis reveals that the CM phase driven by $V_2$ has an
intrinsically fluctuating and partly disordered character,  which is due to
geometric frustration and cannot be fully captured by simple mean-field
approaches.\cite{canocortes:2011fx} The origin of these fluctuations
becomes apparent by considering the atomic limit $t=0$, in which the
honeycomb lattice decouples into two triangular lattices, on each of which
fermions experience the frustrated repulsion $V_2$. Figure~\ref{fig:cmphase}(a)
illustrates the charge distribution on one sublattice. Filled (empty) 
circles correspond to occupied (empty) sites, contributing 1/6 (1/3 when we
consider both sublattices) to the total band filling of 1/2. The remaining
fermions can be distributed over the $\oplus$ sites at an energy cost of
$3V_2$ each. Since all possible configurations have the same classical energy, a highly
degenerate ground state with only partial charge
order results. An analogous situation was analyzed for the Ising model on
the triangular lattice by Wannier.\cite{Wannier:1950hw} Interestingly, 
the energy is also independent of the relative number of occupied $\oplus$
sites on the two sublattices. Consequently, the repulsion $V_2$ is equally
satisfied by configurations with fermions evenly distributed among the two
sublattices, and configurations with ratios 1:2 or 2:1, or anything in
between. In fact, even a nonzero NN repulsion $V_1$ leaves the degeneracy largely
intact (the energy cost for fermions on $\oplus$ sites
becomes $3V_2+V_1$) and does in particular not favor charge imbalance between
the sublattices, as shown in Fig.~\ref{fig:Nk_CM}. The degeneracy
can hence only be lifted by the hopping $t$.

For spinless fermions on the triangular lattice, it has been
shown\cite{hotta2006} that the partial charge order persists also for a
nonzero hopping $t\neq 0$. The extensive degeneracy due to disorder is there reduced to the
three-fold degeneracy of the charge-order pattern, because the additional
fermions form a metal. This metallic yet partially charge-ordered phase was
dubbed a pinball liquid.\cite{hotta2006} Depending on the Hamiltonian, the
electrons not involved in charge order can show
superconductivity~\cite{Morohoshi:2008cm} or topological
order.~\cite{2013arXiv1305.6948K} In the present model, the hopping $t$
connects the two sublattices, and can provide the largest kinetic energy gain
when the densities in the two sublattices are equal. It will consequently
tend to lift the degeneracy between different sublattice occupations in favor
of equal occupancy. Indeed, the charge structure factor $S({\bf q})$ shown in
Fig.~\ref{fig:Nk_CM} does not indicate a charge imbalance between the
sublattices: While $S({\bf q})$ is clearly peaked at the ordering momenta
$\pm K$, the N\'eel signal [corresponding to $l_{\bf q}=2$, see caption of
Fig.~\ref{fig:Nk_CM}] is not enhanced. Its weight and that for other
 momenta ${\bf q} \neq \pm K$ approach a nonzero value for large $V_2$.
In contrast, the structure factor is suppressed to zero for momenta ${\bf q}
\neq \Gamma$ deep in the charge-density-wave phase. These nonzero values of $S(\bm{q})$ for
${\bf q} \neq \pm K$ in the CM phase support the picture that not all fermions
participate in the $K$-modulated charge order. Figure~\ref{fig:Nk_CM} also reveals that a finite (but
moderate) $V_1=2t$ does not increase the N\'eel signal, or indeed induce any
significant changes, similar to the situation on decoupled sublattices.

In contrast to the metallic pinball liquid found in decoupled
sublattices,\cite{hotta2006} the density of states in
Fig.~\ref{fig:cmphase}(b) shows a gap for the CM phase at $V_2=3t$. By
comparing results for different values of $V_2$, we find that, after
initially increasing with $V_2$, the gap saturates deep in the CM phase. [There are
additional high-energy excitations on the scale of $V_2$ outside the energy
range show in Fig.~\ref{fig:cmphase}(b).] For large $V_2$, where the picture
of two coupled pinball liquids is most applicable, the gap becomes independent of
$V_2$ and instead scales with the hopping $t$. The analog of the metallic
pinball liquid in the model~(\ref{eq:modelH1}) is therefore the insulating CM
phase with interaction-independent low-energy excitations at $|\om|\sim t$.

\subsection{Phases at nonzero $V_1$}

\begin{figure}[t] 
  \includegraphics[width=0.425\textwidth]{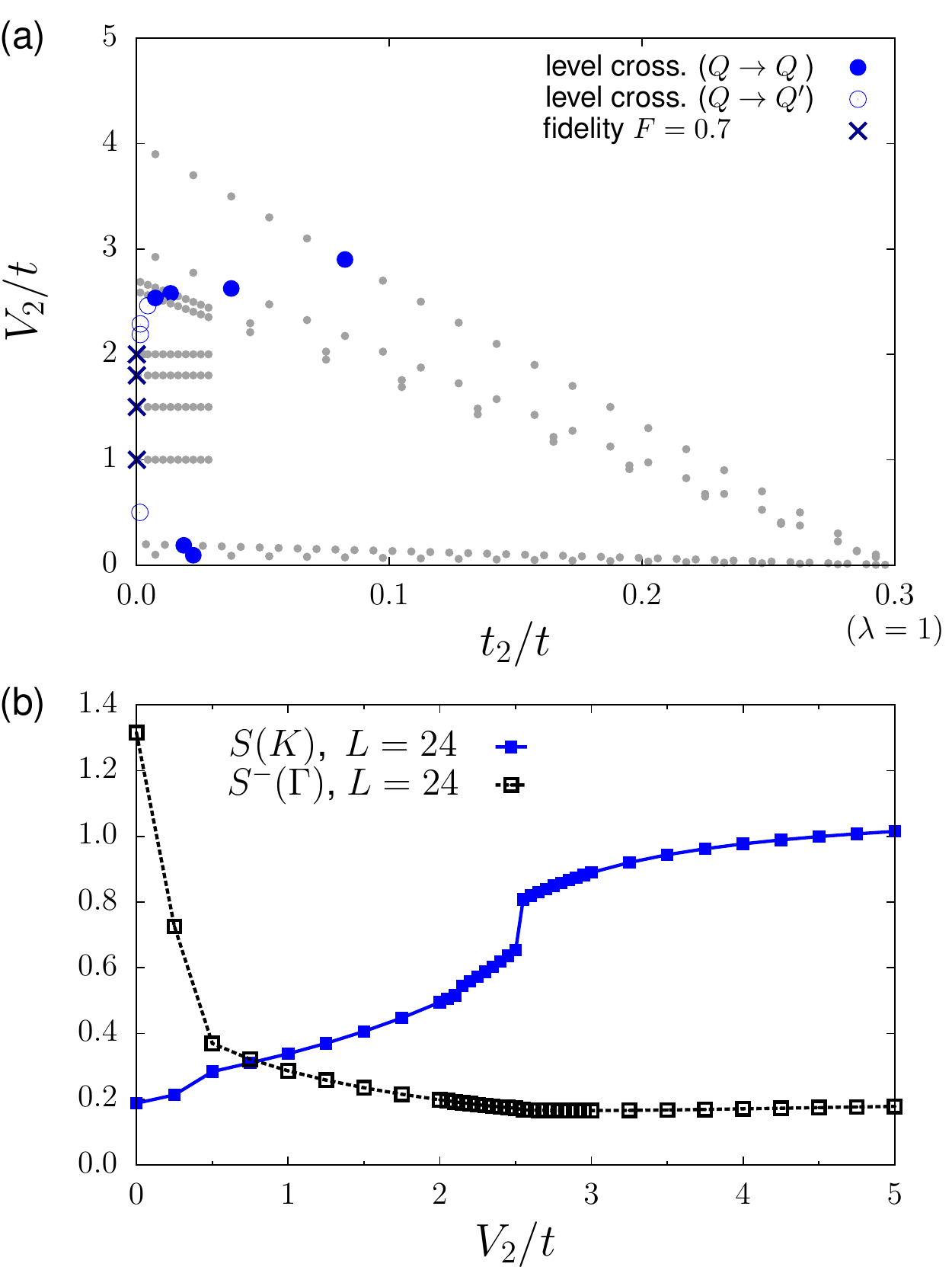}
  \caption{\label{fig:V1} (Color online) 
    (a) As in Fig.~\ref{fig:trans}(b), but for nonzero $V_1=t$. $V_1$ is
    switched on at the same rate as $V_2$, with
    $V_1=t$ along the left vertical axis. All results were obtained from
    exact diagonalization of $\tilde{H}$ using a 24-site cluster.  
    (b) Charge structure factors $S(K)$ and $S^-(\Gamma)$, see
    Eq.~(\ref{eq:Nk}), from exact diagonalization with $L=24$.
  }
\end{figure}

To establish the robustness of our findings at $V_1=0$, we briefly consider
a nonzero $V_1$.  Figure~\ref{fig:V1}(a) shows results similar to
Fig.~\ref{fig:trans}(b), obtained for $\tilde{H}$ by varying the parameters
$t_2$, $V_2$ and $V_1$ along paths that connect the Hamiltonians $\hat{H}_1$ and
$\hat{H}_2$. The starting point in the lower right corner
corresponds to the QAH phase of the Haldane model $\hat{H}_2$ with $t_2=0.3$ and
$V_1=V_2=0$. On the left vertical axis, $t_2=0$, $V_1=t$, and $V_2$ takes on the
values indicated in the plot. (Starting from $\lambda=0$ and while switching
off $t_2$, we switch on $V_1$
and $V_2$ at the same rate.) In addition, we have performed horizontal scans
at fixed $V_1$ and $V_2$ starting from $t_2=0$. 

For paths with $V_2\gtrsim2.5t$ at $\lambda=0$, we find level crossings in
the same ground state momentum sector at finite critical values of $t_2$
(filled circles). As for Fig.~\ref{fig:trans}(b), we interpret these
crossings as quantum phase transitions and hence as the absence of an
adiabatic connection between the ground state of $\hat{H}_1$ and the QAH phase of the Haldane model. Instead,
the results for the charge structure factors in Fig.~\ref{fig:V1}(b) suggest
that the gapped phase at large $V_2$ is again the CM phase. As already 
seen in Fig.~\ref{fig:Nk_CM}, the $V_2$-driven charge modulation is
hardly affected by a small to moderate $V_1$. At intermediate
values $2.1\lesssim V_2/t \lesssim 2.5$, the same cluster-related complications arise as for
Fig.~\ref{fig:trans}(b), namely a level crossing with a change of the ground
state momentum sector. As for $V_1=0$, see Fig.~\ref{fig:trans}(b),
level crossings take place at finite but very small $t_2\approx
0.003t$. The $V_2$-driven charge modulations grow throughout this regime,
see Fig.~\ref{fig:V1}(b), and as argued in Sec.~\ref{sec:CM_V2}, we regard the
level crossings as a finite-size effect rather than indications of an
intermediate phase.  The fact that $V_1$ has a negligible
impact in this parameter regime can be taken as further evidence against
potential intermediate phases that would be either stabilized or destabilized
by a nonzero $V_1$.

For $1\lesssim V_2/t \lesssim 2.1$, the fidelity jumps from $F\approx 1$  to
$F\approx 0.7\approx \sqrt{2}$ when $t_2$ becomes zero, indicating the existence
of a gapless SM phase that is unstable towards the opening of a QAH gap via $t_2$.
The only difference to the case $V_1=0$ shown in
Fig.~\ref{fig:trans}(b) is found at very small $V_2$, where we again
see a level crossing with vanishing fidelity. In agreement with mean-field and numerical
results,\cite{GrushinII,Garciaetal2013} these level crossings indicate the transition
from the QAH phase at $V_1=0$ and $t_2 >0$ to the gapped N\'eel
charge-density-wave state found at $V_1=1, t_2\approx 0, V_2\approx 0$. The
existence of the latter is also supported by the peak in $S^-(\Gamma)$
visible in Fig.~\ref{fig:V1}(b). Between this regime with N\'eel order and the
SM, we again observe a transition regime involving level crossings between different
momentum sectors which we attribute to the cluster geometry.

Except for the vicinity of $V_2=0$, the results for $V_1=t$ are hence very similar
to those for $V_1=0$. Consequently, our conclusion regarding the absence of
an intermediate phase and a direct transition from the SM phase to the CM
phase also holds at nonzero $V_1$ and is hence robust. 

\subsection{Quantum fluctuation effects from cluster perturbation theory}

Our numerical results suggest a direct transition from the Dirac SM to the CM
phase, and hence the absence of an intermediate QAH phase. A better
understanding of this issue can be obtained by systematically adding quantum
fluctuations to the mean-field ansatz using a variant of cluster perturbation
theory.\cite{PhysRevB.66.075129} Within this approach, we treat interactions
and hopping processes inside a finite cluster of $L$ sites
exactly. Single-particle terms that connect different clusters, including the
mean-field decoupled interaction terms, are accounted for in first-order
perturbation theory.\cite{PhysRevB.70.235107}

For the Hamiltonian~(\ref{eq:modelH1}), the mean-field 
decoupling reads $\on_i \on_j \mapsto \las n_i
\ras \on_j + \on_i \las n_j \ras - \las n_i \ras \las n_j \ras
-\las \cdag_i \cnod_j\ras  \cdag_j \cnod_i
-\las \cdag_j \cnod_i\ras  \cdag_i \cnod_j
+|\las \cdag_i \cnod_j\ras|^2$. 
The first three (Hartree) terms can give rise to charge-density-wave order,
whereas the last three (Fock) terms can lead to bond-ordered phases. In
particular, the QAH state emerges from an imaginary bond order parameter
$\chi_{ij} = \las \cdag_i \cnod_j\ras=\pm\rmi|\chi|$ with opposite sign on
the two sublattices.\footnote{A \emph{real} order parameter $\chi_{ij}$ with
opposite sign on the two sublattices likewise opens a gap which is, however,
topologically trivial.}  To explore the most favorable setting for the QAH phase, 
we set the charge-density-wave order parameters $\las n_j \ras$ to zero and
allow for purely imaginary $\chi_{ij}$ only.  For a two-site cluster ($L=2$)
and $V_1=0$, cluster perturbation theory is equivalent to
mean-field theory since \emph{all} $V_2$ interaction terms are decoupled.
Increasing $L$ allows for ordered patterns with a larger unit cell, similar
to mean-field theory,\cite{PhysRevB.81.085105,GrushinII} and additionally
includes short-range quantum fluctuations by treating more and more bonds exactly.

\begin{figure}[t]
  \includegraphics[width=0.425\textwidth,clip]{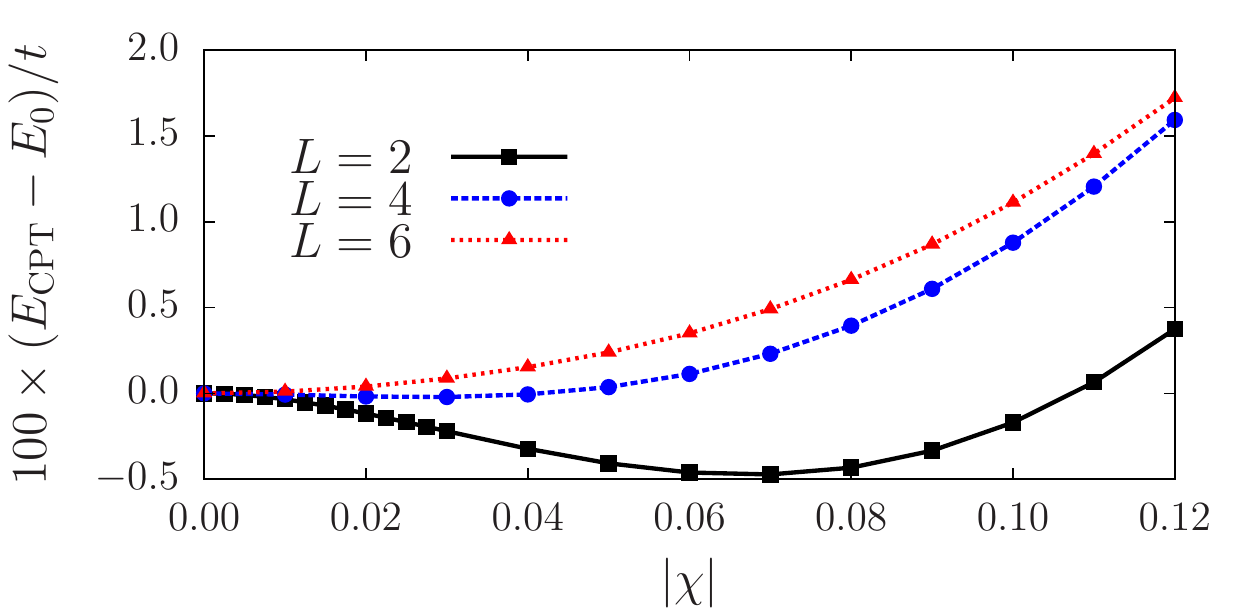}
  \caption{\label{fig:variationalenergy}
    (Color online) Ground-state energy as a function of the bond order
    parameter $\chi_{ij}=\pm\rmi|\chi|$ from cluster perturbation theory
    with different cluster sizes $L$. Here, $V_2=2t$ and
    $V_1=0$. For $L=6$, the energy is minimal at $|\chi|=0$,
    corresponding to the absence of a QAH phase. 
  }
\end{figure}

The impact of quantum fluctuations is apparent from
Fig.~\ref{fig:variationalenergy} which shows the total energy as a function
of $|\chi|$ for clusters with $L=2,4$, and 6 sites. Whereas the QAH state
exists for $V_2=2t$ in mean-field
theory\cite{RaQiHo08,PhysRevB.81.085105,GrushinII} and for $L=2$ in
Fig.~\ref{fig:variationalenergy}, it is quickly suppressed with increasing
$L$. Already for $L=6$, the energy is minimal for $|\chi|=0$, and a QAH phase
is absent.

\subsection{Tendency toward a QAH state at small $V_2$}

\begin{figure}[t]
  \includegraphics[width=0.425\textwidth,clip]{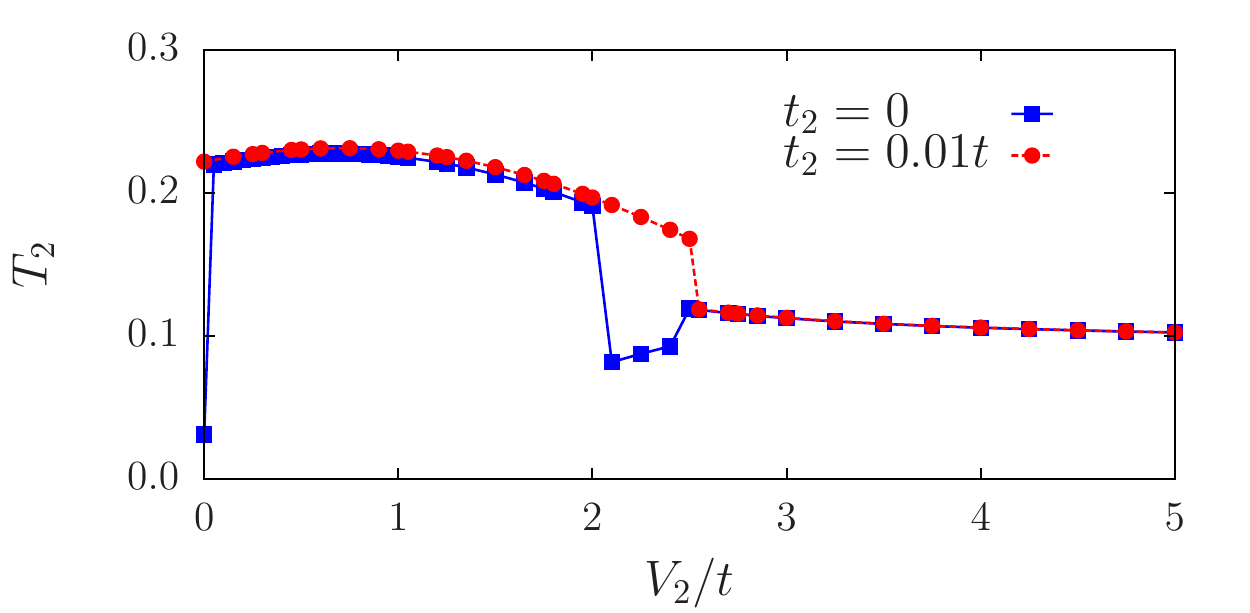}
  \caption{\label{fig:qah_V2}
    (Color online) Susceptibility $T_2$ defined in Eq.~(\ref{eq:T2}) as a
    function of $V_2$. Results were obtained from exact diagonalization of a
    24-site cluster. 
  }
\end{figure}

Further insight into the tendency toward an interaction-induced QAH phase can
be gained from Fig.~\ref{fig:qah_V2}. It shows the susceptibility
\begin{equation}\label{eq:T2}
T_2 =  \frac{1}{L^2}\Big\langle \Big[\sum_{\llas ij \rras} (e^{\rmi \phi_{ij}} c^\dag_i c^\nag_j +
e^{-\rmi \phi_{ij}} c^\dag_j c^\nag_i) \Big]^2\Big\rangle\,,
\end{equation}
related to the Haldane hopping term [see Eq.~(\ref{eq:haldane})], as a
function of $V_2$, as obtained from exact diagonalization. The phase has
been chosen as $\phi_{ij}=\pm\pi/2$. 

We first discuss the case of $t_2=0.01t$, a small symmetry breaking field
that establishes a topological QAH state at $V_2=0$. Upon switching on the
interaction $V_2$, the susceptibility $T_2$ initially increases. At larger
values of $V_2$, the susceptibility decreases with increasing $V_2$, shows
a pronounced drop at around $V_2=2.5t$ and saturates for $V_2\geq 2.5t$.
Essentially the same overall behavior is observed in the absence of a
symmetry-breaking field, \ie, for $t_2=0$. In contrast to $t_2=0.01t$,
the ground state for $t_2=0$ and $V_2=0$ is a semimetal, and $T_2$ is 
much smaller. Around $V_2=2.5t$, we see non-generic signatures related to
the cluster considered, as previously discussed for
Fig.~\ref{fig:phaseszerot2}, which are absent on other clusters and for
$t_2=0.01t$.

The initial increase of $T_2$ with increasing $V_2$ may be interpreted as a
signature of a tendency toward an interaction-driven QAH phase, in agreement with the
fact that a low-energy theory of the model~(\ref{eq:modelH1}) yields an
instability of the SM toward the opening of a topological gap via a staggered
$t_2$ hopping term.\cite{RaQiHo08} Moreover, this increase at weak $V_2$ is likely to
cause the corresponding coupling to increase under a renormalization group
flow. Hence, a weak-coupling stability analysis, similar to the one carried
out for the interaction-generated quantum spin Hall phase in
Ref.~\onlinecite{RaQiHo08}, would likely indicate ordering tendencies toward this
phase. However, the results in Fig.~\ref{fig:qah_V2} reveal a decrease at
larger values of $V_2$, in accordance with the absence of a QAH phase at
$t_2=0$. The absence of such as phase in the model~(\ref{eq:modelH1}),
despite the weak-coupling instability, can be attributed to the vanishing of
the density of states at the Fermi level in the SM phase, which renders the tendency toward
symmetry breaking and spontaneous bond order too weak for a stable
phase to exist. However, the enhancement of Haldane-type bond-order
correlations for small values of $V_2$ suggests that the
balance can be tipped in favor of a QAH phase, so that the latter may be
stabilized in modified or extended models. Weak-coupling instabilities can
occur if the density of states at the Fermi level is finite, for example on
other two-dimensional lattices\cite{PhysRevLett.103.046811,PhysRevB.82.045102,PhysRevB.82.075125}
or in bilayer systems.\cite{PhysRevB.82.205106,PhysRevB.81.041402,PhysRevB.85.235408}

\section{Conclusions}\label{sec:conclusions}

We have revisited the problem of spinless fermions on the honeycomb lattice
with repulsive, nonlocal interactions. Using exact diagonalization, we found
no evidence for the interaction-generated quantum Hall state observed in
previous mean-field treatments of the same
model.\cite{RaQiHo08,PhysRevB.81.085105,GrushinII} Instead, for $V_1=0$, our
data suggest a direct transition from a correlated semimetal to a gapped,
charge-modulated phase at $V_2\approx 2.5t$.

The conclusion regarding the quantum Hall phase is based on the absence of an
adiabatic connection to the ground state of the Haldane model throughout the
gapped parameter region. The instability of the mean-field quantum Hall state
can also be illustrated by including fluctuations around the mean-field
solution in the framework of cluster perturbation theory. At smaller $V_2$,
we found indications for a tendency toward a QAH state that is enhanced by increasing
$V_2$, but the vanishing density of states at the Fermi level limits the
potential energy gain and thereby prevents the formation of a stable QAH phase.
However, the phase may well exist in models with weak-coupling instabilities
related to quadratic band crossing points,\cite{PhysRevLett.103.046811,PhysRevB.82.045102,PhysRevB.82.075125}
where the density of states is finite. A weak-coupling quantum Hall phase
has also been found at the mean-field level in a model of strained graphene.\cite{PhysRevB.88.045425}
To identify the modifications of the model which are necessary for the QAH
phase to exist represents a fascinating topic for future work.

The charge-modulated phase at large $V_2$ turns out to be rather
unconventional and was found to have close relations to frustrated spin
systems and pinball liquids. It is gapped, but the energy of the lowest-lying
excitations becomes independent of the interaction in the strong-coupling
regime and is instead determined by the hopping integral $t$. Hence, while
the phase clearly emerges from a large $V_2$, and therefore is in some sense a
Mott insulator, its band gap is set by the hopping integral $t$, a property
typical of band insulators.

\begin{acknowledgments}%
  We are grateful to F. Assaad, C. Honerkamp, A. Grushin, J. Moore,
  T. Neupert, S. Rachel, A. R\"uegg, J. Venderbos, and M. Vojta for helpful
  discussions, and acknowledge support from the DFG
  Grant No.~Ho~4489/2-1 (FOR 1807) and the Emmy Noether Programme. We also
  thank the authors of Ref.~\onlinecite{Garciaetal2013} for sharing their
  results with us prior to publication. 
\end{acknowledgments}%

\section*{Note added}

During the preparation of this manuscript, we learned about the results of
Ref.~\onlinecite{Garciaetal2013}, in which the model~(\ref{eq:modelH1}) was
studied using exact diagonalization. Both works agree on the absence of an
interaction-generated QAH phase.


%

\end{document}